\documentclass[aps,prc,twocolumn,superscriptaddress,showpacs]{revtex4}
\usepackage[dvips]{graphicx}
\newcommand{\nc}{\newcommand}           
\nc{\vc}[1]	{\mbox{\boldmath $#1$}}	
\nc{\bra}       {\langle}               
\nc{\ket}       {\rangle}               
\nc{\del}       {\partial}               
\nc{\mapleft}[1]{			
 \smash{\mathop{\,			%
  \hbox to 1.5cm{\rightarrowfill}\, }\limits_{#1}}}

\def\JL#1#2#3#4{ {{\rm #1}}\ \textbf{#2}, #4 (#3)}  
\nc{\PR}[3]     {\JL{Phys. Rev.}{#1}{#2}{#3}}
\nc{\PRC}[3]    {\JL{Phys. Rev.~C}{#1}{#2}{#3}}
\nc{\PRA}[3]    {\JL{Phys. Rev.~A}{#1}{#2}{#3}}
\nc{\PRL}[3]    {\JL{Phys. Rev. Lett.}{#1}{#2}{#3}}
\nc{\NP}[3]     {\JL{Nucl. Phys.}{#1}{#2}{#3}}
\nc{\NPA}[3]     {\JL{Nucl. Phys.}{#1}{#2}{#3}}
\nc{\PL}[3]     {\JL{Phys. Lett.}{#1}{#2}{#3}}
\nc{\PLB}[3]    {\JL{Phys. Lett.~B}{#1}{#2}{#3}}
\nc{\PTP}[3]    {\JL{Prog. Theor. Phys.}{#1}{#2}{#3}}
\nc{\PTPS}[3]   {\JL{Prog. Theor. Phys. Suppl.}{#1}{#2}{#3}}
\nc{\PRep}[3]   {\JL{Phys. Rep.}{#1}{#2}{#3}}
\nc{\JP}[3]     {\JL{J. of Phys.}{#1}{#2}{#3}}
\nc{\andvol}[3] {{\it ibid.}\JL{}{#1}{#2}{#3}}

\begin{document}
\title{
Roles of the tensor and pairing correlations on the halo formation in $^{11}$Li
}

\author{Takayuki Myo
\footnote{E-mail address: myo@rcnp.osaka-u.ac.jp}
}
\affiliation{
Research Center for Nuclear Physics (RCNP),
Ibaraki, Osaka 567-0047, Japan}

\author{Kiyoshi Kat\=o
\footnote{E-mail address: kato@nucl.sci.hokudai.ac.jp}
}
\affiliation{
Division of Physics, Graduate School of Science,
Hokkaido University, Sapporo 060-0810, Japan.}

\author{Hiroshi Toki
\footnote{E-mail address: toki@rcnp.osaka-u.ac.jp}
}
\affiliation{
Research Center for Nuclear Physics (RCNP),
Ibaraki, Osaka 567-0047, Japan}

\author{Kiyomi Ikeda
\footnote{E-mail address: k-ikeda@postman.riken.go.jp}
}
\affiliation{
The Institute of Physical and Chemical Research (RIKEN), 
Wako, Saitama 351-0198, Japan.}

\date{\today}

\begin{abstract}
We study the roles of the tensor and pairing correlations on the halo formation in $^{11}$Li with an extended $^9$Li+$n$+$n$ model.
We first solve the ground state of $^9$Li in the shell model basis by taking $2p$-$2h$ states using the Gaussian functions with variational size parameters to take into account the tensor correlation fully. In $^{11}$Li, the tensor and pairing correlations in $^9$Li are Pauli-blocked by additional two neutrons, which work coherently to make the configurations containing the $0p_{1/2}$ state pushed up and close to those containing the $1s_{1/2}$ state. Hence, the pairing interaction works efficiently to mix the two configurations by equal amount and develop the halo structure in $^{11}$Li. For $^{10}$Li, the inversion phenomenon of $s$- and $p$-states is reproduced in the same framework.
Our model furthermore explains the recently observed Coulomb breakup strength and charge radius for $^{11}$Li.
\end{abstract}

\pacs{
21.60.Gx,~
21.10.Pc,~
27.20.+n~
}

\maketitle 

\section{Introduction}
A pioneering secondary-beam experiment found that the size of $^{11}$Li was surprisingly large, which was outside of the common sense of nuclear physics \cite{Ta85}.  This large size was later interpreted as due to the halo structure of two neutrons around the $^9$Li core nucleus \cite{Ha87}.  This finding together with others motivated the nuclear physics community to start a new research field 
for the study of unstable nuclei and built new facilities of Radioactive Ion Beams (RIB) in several laboratories as RIKEN, MSU, GSI, GANIL and others.  Many experimental findings were shown later for $^{11}$Li: a) The halo neutrons have almost equal amount of the $s$-wave component with respect to the $p$-wave component \cite{Si99}.  b) The dipole strength distribution has a large enhancement near the threshold \cite{Na06}.  c) The charge radius is larger than that of $^9$Li \cite{Sa06,Pu06}.

The biggest puzzle from the theory side is the large $s$-wave component for the halo neutrons. If we interpret this fact in the shell model, the shell gap at $N=8$ has to disappear. However, the mean field treatment of a central force is not able to provide the disappearance of the $N=8$ shell gap. So far, there were many theoretical studies for $^{11}$Li\cite{To90b,Es92,To94,De97,Ba97,Mu98,Va02,My02,Ga02a,Ga02b,Mi06,Br06} and essentially all the theoretical works of $^{11}$Li had to accept that the $1s_{1/2}$ single particle state is brought down to the degenerated energy position with the $0p_{1/2}$ state without knowing its reason\cite{To94}. It is therefore the real challenge for theoretician to understand this
disappearance of the $N=8$ shell gap, to be called the $s$-$p$ shell gap problem, which is worked out in this paper by developing a framework to treat the tensor force explicitly in the nucleon-nucleon interaction. The halo structure of $^{11}$Li is also related with the $1s$-state and the $0p$-state in $^{10}$Li.  Several experiments suggest that the dual states of the $s_{1/2}$-state coupled to the $3/2^-$ proton state appear close to the threshold energy of $^9$Li+$n$ together with the dual states of the $p_{1/2}$-state \cite{Th99,Cha01,Je06}.
This property is known as the inversion problem of $s$- and $p$-states seen in the $N=7$ isotone\cite{Cha01}.

The pairing correlation was considered important for the $s$-$p$ shell gap problem\cite{Es92,My02}. The additional neutrons would act to block the pairing correlation of the core when one neutron in $^{10}$Li or two neutrons in $^{11}$Li are added in the $p$-orbit\cite{My02,Sa93,Ka99}.
Our calculations with this pairing-blocking effect improve somewhat the descriptions of $^{10}$Li and $^{11}$Li, but it was not sufficient to explain the large $s$-wave component in $^{11}$Li\cite{My02}. We also pointed out the different roles of the pairing-blocking between $^{10}$Li and $^{11}$Li. We need further mechanism for the increase of the $s$-wave component in $^{11}$Li.

The tensor force, on the other hand, plays an important role in the nuclear structure.  For example, the contribution of the tensor force in the binding of $^4$He is comparable to that of the central force\cite{Ak86,Ka01}.  The tensor correlation induced by the tensor force was demonstrated to be important for the $^4$He+$n$ system \cite{Te60,Na59,My05}. In our recent study \cite{My07}, we developed a theoretical framework of the tensor-optimized shell model to treat the tensor force in the shell model basis explicitly including $2p$-$2h$ excitations.
We found that the $(0s_{1/2})^{-2}(0p_{1/2})^2$ excitation of proton-neutron pair has a special importance in describing the tensor correlation in $^4$He \cite{Su04,My05,My07}. In the $^4$He+$n$ system, because this $2p$-$2h$ excitation receives the strong Pauli-blocking from the last neutron occupying the $p_{1/2}$-orbit, a considerable amount of the $p_{1/2}$-$p_{3/2}$ splitting energy in $^5$He is reproduced \cite{My05}. This Pauli-blocking effect from the $p_{1/2}$-orbit caused by the tensor force should be present also for $^{11}$Li.

Hence, it is very interesting to study the effect of the tensor correlation together with the pairing correlation for the $s$-$p$ shell gap problem in $^{11}$Li.  This is the purpose of this paper. To this end, we shall perform the configuration mixing based on the shell model framework for $^9$Li to describe the tensor and pairing correlations explicitly. In particular, we pay attention to the special features of the tensor correlation. For $^{11}$Li, we shall solve the configuration mixing of the $^9$Li+$n$+$n$ problem which treats both correlations, and investigate further the Coulomb breakup strength of $^{11}$Li and other observables to see the effect of these correlations.
We also investigate the inversion phenomena of $^{10}$Li considering the tensor and pairing correlations.

\section{Model}

\subsection{Coupled $^9$Li+n+n model of $^{11}$Li}

We shall begin with the introduction of the model for $^9$Li, whose Hamiltonian is given as
\begin{eqnarray}
    H(\mbox{$^9$Li})
&=& \sum_{i=1}^9{t_i} - t_G  + \sum_{i<j} v_{ij}\ .
    \label{H9}
\end{eqnarray}
Here, $t_i$, $t_G$, and $v_{ij}$ are the kinetic energy of each nucleon, the center-of-mass (c.m.) term and the two-body $NN$ interaction consisting of central, spin-orbit, tensor and Coulomb terms, respectively. The wave function of $^9$Li($3/2^-$) is described in the tensor-optimized shell model\cite{My05,My07}. We express $^9$Li by a multi-configuration,
\begin{eqnarray}
  \Psi(^{9}\mbox{Li})=\sum_i^N a_i\, \Phi^{3/2^-}_i,
  \label{WF9}
\end{eqnarray}
where we consider up to the $2p$-$2h$ excitations within the $0p$ shell for $\Phi^{3/2^-}_i$ in a shell model type wave function, and
$N$ is the configuration number. Based on the previous study of the tensor-optimized shell model\cite{My05,My07}, we adopt the spatially modified harmonic oscillator wave function (Gaussian function) as a single particle orbit and treat the length parameters $b_\alpha$ of every orbit $\alpha$ of $0s$, $0p_{1/2}$ and $0p_{3/2}$ as variational parameters. This variation is shown to be important to optimize the tensor correlation\cite{My05,My07,Su04,Og05}.

Following the procedure of the tensor-optimized shell model, we solve the variational equation for the Hamiltonian of $^9$Li and determine $\{a_i\}$ in Eq.~(\ref{WF9}) and the length parameters $\{b_{\alpha}\}$ of three orbits. The variation of the energy expectation value with respect to the total wave function $\Psi(^9{\rm Li})$ is given by
\begin{eqnarray}
\delta\frac{\bra\Psi|H(\mbox{$^9$Li})|\Psi\ket}{\bra\Psi|\Psi\ket}&=&0\ ,
\end{eqnarray}
which leads to the following equations:
\begin{eqnarray}
    \frac{\del \bra\Psi| H(\mbox{$^9$Li}) - E |\Psi \ket} {\del b_{\alpha}}
=   0,~~
    \frac{\del \bra\Psi| H(\mbox{$^9$Li}) - E |\Psi \ket} {\del a_{i}}
=   0.
\end{eqnarray}
Here, $E$ is the total energy of $^9$Li. The parameters $\{b_{\alpha}\}$ for the Gaussian bases appear in non-linear forms in the total energy $E$. We solve two kinds of variational equations in the following steps. First, fixing all the length parameters $b_{\alpha}$, we solve the linear equation for $\{a_{i}\}$ as an eigenvalue problem for $H$($^9$Li). We thereby obtain the eigenvalue $E$, which is a function of $\{b_{\alpha}\}$. Next, we try various sets of the length parameters $\{b_{\alpha}\}$ to find the solution which minimizes the energy of $^9$Li. In this wave function, we can optimize the radial form of single particle orbit appropriately so as to describe the spatial shrinkage of the particle state, which is important for the tensor correlation\cite{My05,My07,Su04,Og05}.

For $^{11}$Li and $^{10}$Li, their Hamiltonians are written in terms of $^9$Li+$n$+$n$ and $^9$Li+$n$, respectively, and are given as
\begin{eqnarray}
  H(\mbox{$^{11}$Li})
&=&H(\mbox{$^9$Li})
+   \sum_{k=0}^2{T_k} - T^{(3)}_G
+   \sum_{k=1}^2V_{cn,k}
\nonumber\\
&& + V_{nn}, 
    \label{H11}
    \\
  H(\mbox{$^{10}$Li})
&=&H(\mbox{$^9$Li})
+   \sum_{k=0}^1{T_k} - T^{(2)}_G + V_{cn},
    \label{H10}
\end{eqnarray}
where $H(\mbox{$^9$Li})$, $T_k$, $T^{(3)}_G$ and $T^{(2)}_G$  are the internal Hamiltonian of $^9$Li given by Eq.~(\ref{H9}), the kinetic energies of each cluster ($k=0$ for $^9$Li) and the c.m. terms of three or two cluster systems, respectively. ${V}_{cn,k}$ are the $^9$Li core-$n$ interaction ($k=1,2$) and $V_{nn}$ is the interaction between last two neutrons. The wave functions of $^{11}$Li and $^{10}$Li with the spin $J$ and $J^\prime$, respectively, are given as
\begin{eqnarray}
    \Psi^J(^{11}{\rm Li})
&=& \sum_i^N{\cal A}\left\{ [\Phi^{3/2^-}_i, \chi^{J_0}_i(nn)]^J \right\},
    \label{WF11}
    \\
    \Psi^{J^\prime}(^{10}{\rm Li})
&=& \sum_i^N{\cal A}\left\{ [\Phi^{3/2^-}_i, \chi^{J^\prime_0}_i(n)]^{J^\prime} \right\}.
    \label{WF10}
\end{eqnarray}
We obtain the coupled differential equations for the neutron wave functions $\chi^{J_0}(nn)$ and $\chi^{J^\prime_0}(n)$, where $J_0$ and $J_0^\prime$ are the spins of the additional neutron part of $^{11}$Li and $^{10}$Li, respectively. To obtain the total wave function $\Psi^J(^{11}{\rm Li})$ and $\Psi^{J^\prime}(^{10}{\rm Li})$, we actually use the orthogonality condition model (OCM) \cite{To90b,My02,Ao06} to treat the antisymmetrization between last neutrons and neutrons in $^9$Li. In OCM, the neutron wave functions $\chi$ are imposed to be orthogonal to the occupied orbits by neutrons in $^9$Li, which depend on the configuration $\Phi^{3/2^-}_i$ in Eq.~(\ref{WF9}). We obtain the following coupled Schr\"odinger equations with OCM for the set of the wave functions $\{\chi_i^{J_0}(nn)\}$ for $^{11}$Li and $\{\chi_i^{J^\prime_0}(n)\}$ for $^{10}$Li, where $i=1,\cdots,N$:
\begin{eqnarray}
\lefteqn{\hspace*{-3.cm}
\sum_{j=1}^N \left[ \left(T_{\rm rel}^{(3)} + \sum_{k=1}^2V_{cn,k}+V_{nn}+ \Lambda_i \right) \delta_{ij} + h_{ij}(^9{\rm Li})\right]
}
\nonumber
\\
\times~\chi_j^{J_0}(nn)&=&E\ \chi_i^{J_0}(nn),
\label{OCM11}
\\
\lefteqn{\hspace*{-3.cm}
\sum_{j=1}^N \left[ \left(T_{\rm rel}^{(2)} + V_{cn} + \Lambda_i \right) \delta_{ij} + h_{ij}(^9{\rm Li})\right]}
\nonumber
\\
\times~\chi_j^{J_0^\prime}(n)&=&E\ \chi_i^{J_0^\prime}(n),
\label{OCM10}
\\
\Lambda_i
&=& \lambda \sum_{\alpha\in \Phi_i(^{9}{\rm Li})} |\psi_\alpha \ket \bra \psi_\alpha|,
\end{eqnarray}
where $h_{ij}(^9{\rm Li})=\bra \Phi_i^{3/2^-} | H(^{9}{\rm Li}) | \Phi_j^{3/2^-} \ket$. $T^{(3)}_{\rm rel}$ and $T^{(2)}_{\rm rel}$ are the total kinetic energies consisting of the relative motions for $^{11}$Li and $^{10}$Li, respectively. $\Lambda_i$ is the projection operator to remove the Pauli forbidden states $\psi_\alpha$ from the relative wave functions\cite{Ku86,Ka99}, where $\psi_\alpha$ is the occupied single particle wave function of the orbit $\alpha$ in $^9$Li. This $\Lambda_i$ depends on the neutron occupied orbits in the configuration $\Phi^{3/2^-}_i$ of $^{9}$Li 
and plays an essential role to produce the Pauli-blocking in $^{11}$Li and $^{10}$Li, explained later.
The value of $\lambda$ is taken large as $10^6$~MeV in the present calculation in order to project out the components of the Pauli forbidden states into an unphysical energy region.
Here, we keep the length parameters \{$b_\alpha$\} of the single particle wave functions as those obtained for $^9$Li. 

We explain the method of treating the orthogonality condition including the particle-hole excitations of $^9$Li in more detail\cite{My02,My05}. 
When the neutron orbit in $^9$Li is fully occupied, the orthogonality condition for the last neutrons to this orbit is given by $\Lambda_i$ in Eqs.~(\ref{OCM11}) and (\ref{OCM10}). When neutron orbits in $^9$Li are partially occupied, such as in the $2p$-2$h$ states, 
the last neutrons can occupy these orbits with particular probabilities, which are determined by the fractional parentage coefficients 
of the total wave functions of $^{10,11}$Li consisting of $^9$Li and the last neutrons.

We describe the two neutron wave functions $\chi$ in Eq.~(\ref{OCM11}) for $^{11}$Li precisely in a few-body approach of the hybrid-TV model\cite{To90b,My02,Ao95};
\begin{eqnarray}
      \chi^{J_0}_i(nn)
&=&   \chi^{J_0}_i(nn,\vc{\xi}_V)+\chi^{J_0}_{i}(nn,\vc{\xi}_T),
      \label{TV}
\end{eqnarray}
where $\vc{\xi}_V$ and $\vc{\xi}_T$ are V-type and T-type coordinate sets of the three-body system, respectively. The radial part of the relative wave function is expanded with a finite number of Gaussian basis functions centered at the origin. We use at most 15 Gaussian basis functions with the maximum range parameter 30 fm to describe the loosely bound wave function of neutron halo\cite{Ao06}.

Here, we discuss the coupling between $^9$Li and the last neutrons, whose details were already explained in the pairing-blocking case\cite{Ka99,My02,My03}. We consider the case of $^{11}$Li. In the present three-body model, the Pauli forbidden states for the relative motion 
provides the Pauli-blocking effect caused by the last two neutrons\cite{My02,Ka99}. This blocking depends on the relative distance between $^9${Li} and the two neutrons, and change the structure of $^9$Li inside $^{11}$Li, which is determined variationally to minimize the energy of the $^{11}$Li ground state. Asymptotically, when the last two neutrons are far away from $^9${Li} ($\vc{\xi}_{V,T}\to\infty$), the effects of antisymmetrization and the interaction between $^9$Li and two neutrons vanish in Eq.~(\ref{OCM11}). Therefore, any coupling between $^9$Li and two neutrons disappears and $^9$Li becomes its ground state. Namely, the mixing coefficients $\{a_i\}$ are the same as those obtained in Eq.~(\ref{WF9}):
\begin{eqnarray}
	\Phi^J(^{11}{\rm Li})
\mapleft{\vc{\xi}_{V,T}\to\infty}
	\left[ \Psi(^{9}\mbox{Li}), \chi^{J_0}(nn) \right]^J,
	\label{asympt}
	\\
 \Psi(^{9}\mbox{Li})=\sum_i^N a_i\, \Phi^{3/2^-}_i.
\end{eqnarray}
Therefore, it is easy to obtain the following asymptotic forms of $\{\chi^{J_0}_i(nn)\}$ from the above two relations:
\begin{eqnarray}
    \chi^{J_0}_i(nn)
&\mapleft{\vc{\xi}_{V,T}\to\infty} & a_i \cdot \chi^{J_0}(nn),
    \label{asympt2}
\end{eqnarray}
where $i=1,\cdots,N$. Eq.~(\ref{asympt2}) implies that the asymptotic wave function of two neutrons $\chi^{J_0}_i(nn)$ is decomposed into the internal amplitude $a_i$ of $^9$Li and the relative wave function $\chi^{J_0}(nn)$. Eqs.(\ref{asympt})-(\ref{asympt2}) give the boundary condition of the present coupled three-body model of $^{11}$Li. Contrastingly, when the two neutrons are close to $^9$Li, the two neutrons dynamically couple to the configuration $\Phi^{3/2^-}_i$ of $^9$Li satisfying the Pauli principle. This coupling changes $\{a_i\}$ of $^9$Li from those of the $^9${Li} ground state, and makes the tensor and pairing correlations to be different from those in the isolated case.
For $^{10}$Li, the similar coupling scheme is considered. The dynamical effect of the coupling arising from the Pauli-blocking is explained in the results in detail.

\subsection{Effective interactions}

We explain here the interactions employed in Hamiltonians in Eqs.~(\ref{H9}), (\ref{H11}) and (\ref{H10}). Before explaining the present interactions, we give a brief review of the situation of the treatment of the effective interactions for the study of $^{9,10,11}$Li. As was mentioned, most theoretical studies based on the three-body model of $^{11}$Li employ the state-dependent $^9$Li-$n$ potential where only the $s$-wave potential is made deeper than other partial waves\cite{To94}, while the $^9$Li core is described as inert.
This state-dependence in the $^9$Li-$n$ potential is phenomenologically determined in order to satisfy the experimental observations of a large $s^2$ component and a two-neutron-separation energy of $^{11}$Li, and a virtual $s$-state in $^{10}$Li, simultaneously. On the other hand, for the $nn$ part, the interaction having a mild short-range repulsion\cite{Ba97,Br06} or the density-dependent one are often used\cite{Es92}. However, even in the microscopic cluster models using an unique effective $NN$ interaction consisting of the central and $LS$ forces\cite{De97,Va02}, the $s$-$p$ shell gap problem in $^{11}$Li and $^{10}$Li cannot be solved simultaneously.
From these results, we consider that the usual approach based on the effective central and $LS$ interactions may be insufficient to explain the exotic structures of $^{10,11}$Li. For this problem, even the so-called ab-initio calculations using the realistic $NN$ interactions, such as Green's function Monte Calro\cite{Pi04}, do not provide good results for $^{11}$Li. 

In this study, we focus on the tensor correlation, which is newly considered to figure out the $s$-$p$ shell gap problem. To do this, we extend the three-body model of $^{11}$Li to incorporate the tensor correlation fully, in particular, for the $^9$Li part. In the present study, our policy for the study of $^{11}$Li is to use the experimental informations and the corresponding theoretical knowledge for $^9$Li and $^{10}$Li as much as possible. Following this policy, we explain our interactions in three terms; $v_{ij}$ of $H(^9{\rm Li})$ in Eq.~(\ref{H9}), core-$n$ $V_{cn}$ and $n$-$n$ $V_{nn}$ of the Hamiltonians in Eqs.~(\ref{H11}) and (\ref{H10}). 

For the potential $V_{nn}$ between the last two neutrons, we take a realistic interaction AV8$^\prime$ in Eq.~(\ref{H11}). Our interest is to see the $n$-$n$ correlation in the two-neutron halo structure, and therefore it is necessary to solve two-neutron relative motion without any assumption. For this purpose, our model space of two neutrons using the hybrid -TV model shown in Eq.~(\ref{TV}) has no restriction and wide enough to describe the short range correlation under the realistic nuclear interaction AV8$^\prime$. Therefore, there is no parameter in the potential $V_{nn}$.  

The $^9$Li-$n$ potential, $V_{c n}$, in Eqs.~(\ref{H11}) and (\ref{H10}) is given by folding an effective interaction, the MHN interaction\cite{Ha71,Fu80}, which is obtained by the $G$-matrix calculation and frequently used in the cluster study of light nuclei \cite{To90b,Ka99,Ao06,Fu80,Ao02}. In the $^9$Li+$n$ system, the folding potential for the $^9$Li density calculated by using H.O. wave function has been discussed to reproduce the proper energies of the $^{10}$Li spectra\cite{To90b,Ka99,My02}. Furthermore, considering the small one-neutron-separation energy of $^9$Li and a long-range exponential tail of the density, we improve the tail behavior of the folding potential to have a Yukawa type form\cite{My02,My03}. Any state-dependence is not used in the present $^9$Li-$n$ potential, such as a deeper potential for the $s$-wave.  This is possible because the Pauli blocking effect of the single particle state is in action and the state with the $p_{1/2}$ orbit is pushed up in energy and becomes close to the state with the $s_{1/2}$ state\cite{Ka99,My02}. We will discuss the results on $^{10}$Li after the discussion on $^{11}$Li. We introduce one parameter, $\delta$, which is the second-range strength of the MHN potential in the calculation of the $^9$Li-$n$ potential to describe the starting energy dependence dominantly coming from the tensor force in the $G$-matrix calculation\cite{Fu80,Ao06}. In the present calculation, we chose this $\delta$ parameter to reproduce the two-neutron-separation energy of $^{11}$Li as 0.31 MeV after working out the tensor and the pairing correlation effects as explained later. It is found that this folding potential also reproduces the positions of the $p$-wave resonances in $^{10}$Li, just above the $^9$Li+$n$ threshold energy\cite{Th99}, as shown in the results.

Now we discuss the choice of the interaction between nucleons in the $^9$Li core; $v_{ij}$ in $H(^9{\rm Li})$, where we use the shell model wave functions for the $^9$Li core in Eq.~(\ref{WF9}). Since our main interest in this work is to investigate the role of the tensor force on the two-neutron halo formation, we describe the tensor correlation in addition to the pairing correlation in the $^9$Li core based on the policy mentioned above. Along this line, recently we have many interesting works\cite{Su04,Og05,To02,Ak04,Ik04}. We have also studied the role of the tensor force in the shell model framework, and proposed the tensor-optimized shell model.\cite{My05,My07} As a reliable effective interaction considered from those studies, in this calculation, we use GA proposed by Akaishi\cite{My05,Ak04,Ik04} for $v_{ij}$ in Eqs.~(\ref{H9}), (\ref{H11}) and (\ref{H10}). This effective interaction GA has a term of the tensor force obtained from the $G$-matrix calculation using the AV8$^\prime$ realistic potential keeping the large momentum space\cite{Ak04,Ik04}. In GA, the obtained $^9$Li wave function in Eq.~(\ref{WF9}) shows smaller matter radius than the observed one 
due to the high momentum component produced by the tensor correlation\cite{My05,Su04,Og05}.
Hence, we have to adjust the central force, which is done by changing the second range of the central force by reducing the strength by $21.5\%$ and increasing the range by 0.185 fm to reproduce the observed binding energy and the matter radius of $^9$Li in the same manner as done for $^4$He\cite{My05,My07}.

\subsection{Tensor correlation in the Gaussian expansion method}

\begin{figure}[t]
\centering
\includegraphics[width=7.5cm,clip]{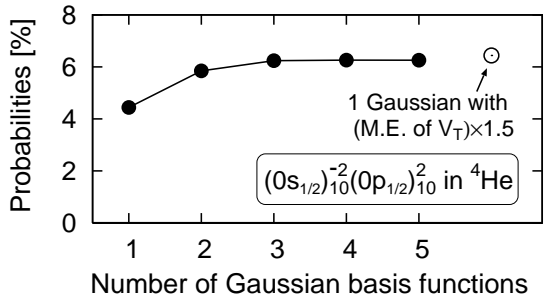}
\caption{Probabilities of the $(0s_{1/2})^{-2}_{10}(0p_{1/2})^{2}_{10}$ component in $^4$He in the Gaussian expansion method are shown as a function of the number of Gaussian basis. Two subscripts 10 represents spin and isospin for the two-nucleon pair, respectively.
The converged value is reproduced by enhancing the tensor matrix elements with one Gaussian basis by 50\%.}
\label{4He}
\end{figure}

In the description of the tensor correlation, in principle, we can work out a large space to include the full effect of the tensor force by taking $2p$-$2h$ states with very high angular momenta\cite{My07}. In order to avoid large computational efforts without loss of the physical importance in the result, we restrict the $2p$-$2h$ shell model states within the $p$-wave states for the description of $^9$Li with a single Gaussian basis. We have studied that the superposition of the Gaussian bases improves the description of the spatial shrinkage for the particle states caused by the tensor correlation\cite{My05,My07}. In this case, the wave function of the particle state $\psi_\alpha$ in $^9$Li is expanded with a finite number of Gaussian basis functions in a $jj$ coupling scheme as
\begin{eqnarray}
    \psi_\alpha
&=& \sum_{n=1}^{N_\alpha} C_{\alpha,n}\ \phi_{\alpha}^n(\vc{r},b_{\alpha,n}),
    \\
    \phi_{\alpha}^n(\vc{r},b_{\alpha,n})
&=& {\cal N}_{\alpha,n} r^{l_\alpha} e^{-(r/b_{\alpha,n})^2/2} [Y_{l_\alpha}(\hat{\vc{r}}),\chi^\sigma_{1/2}]_{j_\alpha}.
\end{eqnarray}
Here $n$ is an index for the Gaussian basis with the length parameter $b_{\alpha,n}$. A basis number and the normalization factor for the basis are given by $N_\alpha$ and ${\cal N}_{\alpha,n}$, respectively. The coefficients $\{C_{\alpha,n}\}$ are determined variationally for the total wave function of $^9$Li in Eq.~(\ref{WF9}). Using this method, so-called the Gaussian expansion method (GEM)\cite{Hi03},
the wave functions of the particle states are improved with an appropriate radial form, where the set of $\{b_{\alpha,n}\}$ is suitably chosen\cite{Hi03,My07}. In particular, it was shown that the particle-hole excitations induced by the tensor force increase\cite{My07}. We have confirmed the GEM effect on the $(0s_{1/2})^{-2}(0p_{1/2})^{2}$ component for $^4$He in Fig.~\ref{4He}, since the similar GEM effect is expected for $^9$Li. As the number of Gaussian basis increases for the particle states, the $(0s_{1/2})^{-2}(0p_{1/2})^{2}$ component increases and converges with three Gaussians. This converged value could be reproduced by increasing the matrix elements of the tensor force with a single Gaussian basis by 50\% as shown in Fig.~\ref{4He}.

Similarly, the GEM effect also affects the Pauli-blocking caused by adding a neutron into the occupied neutron orbit in the core. 
In the scattering problem of the $^4$He+$n$ system, we checked that the GEM effect on the Pauli-blocking is reproduced using the enhanced tensor matrix elements with a single Gaussian basis\cite{My05,My05b,Ik06}. 
The Pauli-blocking could be considered to be almost proportional to the overlap between the wave functions of neutrons inside and outside the core.
In this sense, the single particle properties of the particle states of the core can be described using the enhanced tensor matrix elements.
Therefore, in the present study, we adopt this enhanced tensor matrix elements with a single Gaussian basis in order to simulate the GEM effect.

\section{Results}

\subsection{$^9$Li}

\begin{figure}[t]
\centering
\includegraphics[width=8.2cm,clip]{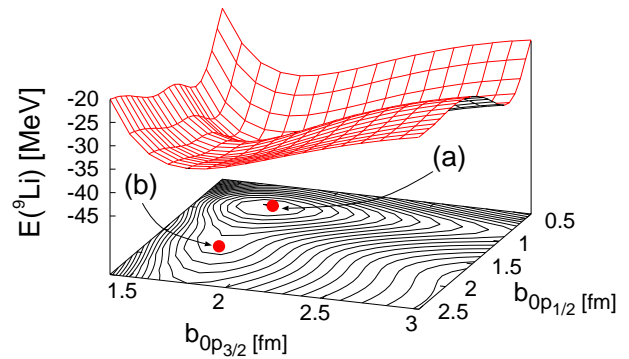}
\caption{(Color online) Energy surface of $^9$Li with respect to the length parameters {$b_\alpha$} of $0p$ orbits.  The two minima indicated by (a) and (b) in the contour map correspond to the states due to the tensor correlation and the paring correlation, respectively.
}
\label{9Li_ene}
\end{figure}

We first show the results of the $^9$Li properties, which give a dynamical influence on the motion of last neutrons above the $^9$Li core
in $^{11,10}$Li. In Fig.~\ref{9Li_ene}, we display the energy surface of $^9$Li as functions of the length parameters of two $0p$ orbits, where $b_{0s}$ is already optimized as 1.45 fm. There are two energy minima, (a) and (b), which have almost a common $b_{0p_{3/2}}$ value of 1.7-1.8 fm, and a small (0.85 fm) and a large (1.8 fm) $b_{0p_{1/2}}$ values, respectively. The properties of two minima are listed in Table \ref{tab:9Li} with the dominant $2p$-$2h$ configurations and their probabilities. It is found that the minimum (a) shows a large tensor contribution, while the minimum (b) does not. Among the $2p$-$2h$ configurations, the largest probabilities are given by $(0s)^{-2}_{10}(0p_{1/2})^2_{10}$ for (a), similar to the results in Ref.~\cite{My05,My07}, and $(0p_{3/2})^{-2}_{01}(0p_{1/2})^{2}_{01}$, namely the $0p$ shell pairing correlation for (b). These results indicate that the minima (a) and (b) represent the different correlations of the tensor and pairing characters, respectively. The spatial properties are also different from each other; the tensor correlation is optimized with spatially shrunk excited nucleons for (a) and the pairing correlation is optimized when two $0p$ orbits make a large spatial overlap for (b). In Table~\ref{tab:9Li}, we show the results of the superposition of minima (a) and (b), named as (c), to obtain a $^9$Li wave function including the tensor and pairing correlations, simultaneously. For (c), the favored two configurations in each minimum (a) and (b) are still mixed with the $0p$-$0h$ one, and the property of the tensor correlation is kept in (c). The superposed $^9$Li wave function possesses both the tensor and pairing correlations.

\begin{table}[t]
\caption{Properties of $^9$Li with configuration mixing.}
\label{tab:9Li}
\begin{center}
\begin{ruledtabular}
\begin{tabular}{c|ccccc}
                             & \multicolumn{3}{c}{Present}  & Expt. \\
                             &   (a)   &  (b)     &  (c)    &       \\
\noalign{\hrule height 0.5pt}
 E [MeV]                     & $-43.8$ & $-37.3$  & $-45.3$ &  $-45.3$\\
 $\langle V_T\rangle $ [MeV] & $-22.6$ &~~$-1.8$  & $-20.7$ &  ---  \\
\noalign{\hrule height 0.5pt}
 $R_m$                 [fm]  &  2.30   & 2.32     & $2.31$  &   2.32$\pm$0.02\cite{Ta88b} \\
\noalign{\hrule height 0.5pt}
 $0p$-$0h$                                         & 91.2  &  60.1 & $82.9$  & --- \\
 $(0p_{3/2})^{-2}_{01}(0p_{1/2})^2_{01}$           & 0.03  &  37.1 & ~$9.0$  & --- \\
 $(0s_{1/2})^{-2}_{10}      (0p_{1/2})^2_{10}$           & 8.2   &   1.8 & ~$7.2$  & --- \\
\end{tabular}
\end{ruledtabular}
\end{center}
\end{table}

\begin{figure}[t]
\centering
\includegraphics[width=8.2cm,clip]{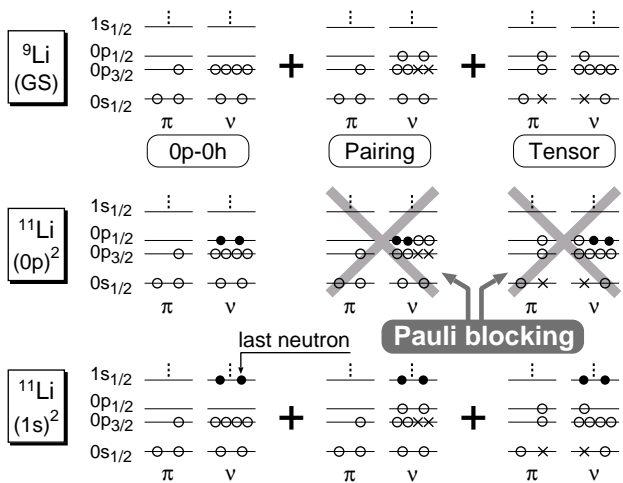}
\caption{Schematic illustration for the Pauli-blocking in $^{11}$Li.
Details are described in the text.}
\label{Pauli}
\end{figure}

\subsection{Pauli-blocking effect in $^{11}$Li}

We discuss here the Pauli-blocking effect in $^{11}$Li and $^{10}$Li. 
We mainly consider the case of $^{11}$Li, as shown in Fig.~\ref{Pauli}.
For the $^9$Li ground state (GS, upper panel), in addition to the $0p$-$0h$ state, $2p$-$2h$ states caused by the tensor and pairing correlations are strongly mixed. 
Let us add two neutrons more to $^9$Li.
When two neutrons occupy the $0p_{1/2}$-orbit (middle panel), 
the $2p$-$2h$ excitations of the tensor and pairing correlations in $^9$Li are Pauli-blocked, simultaneously\cite{My02}.
Accordingly, the correlation energy of $^9$Li is partially lost inside $^{11}$Li.
For the $(1s)^2$ case of two neutrons (lower panel),
the Pauli-blocking does not occur and $^9$Li gains its correlation energy fully by the configuration mixing with the $2p$-$2h$ excitations.
Hence, the relative energy distance between $(0p)^2$ and $(1s)^2$ configurations of $^{11}$Li is expected 
to become small to break the magicity in $^{11}$Li.
The same effect is also expected to explain the inversion phenomena of $1s$- and $p$-states in $^{10}$Li.

In order to confirm the above expectation of the blocking effect on the $(1s)^2$ configuration of $^{11}$Li, we discuss here the configuration mixing including $sd$-shell for $^9$Li. The $(0p_{3/2})^{-2}_{01}(1s)^2_{01}$ neutron pairing excitation in $^9$Li is negligible\cite{Ka99}, and the probability of the $(0s)^{-2}_{10}[(1s)(0d_{3/2})]_{10}$ excitation induced by the tensor force is around $2\%$\cite{My07}. The latter excitation is a proton-neutron pair, in which the $1s$-state is spatially shrunk about a half size of $b_{0s}$ due to the tensor correlation\cite{My07}. When the $1s$-state is occupied by a neutron in $^9$Li, this $1s$-state brings a small overlap with the spatially extended $1s$-orbit of the last neutrons in $^{11}$Li. Then we have estimated that the blocking effect on the $(1s)^2$ component of $^{11}$Li for this proton-neutron excitation is very small. We consider that the characteristics of the blocking effect for $^{11}$Li would not change, even if we include the $sd$-shell for $^9$Li and $^{11}$Li.

\subsection{$^{11}$Li}

We perform the coupled three-body calculation of $^{11}$Li considering the tensor and pairing correlations fully, named as ``Present''. In order to see the individual effects of the tensor and paring correlations, we also compare the results with other three kinds of calculations for $^{11}$Li with different descriptions of $^9$Li.
``Inert core'' is only the $0p$-$0h$ configuration of $^9$Li. 
``Tensor'' and ``Pairing'' are the ones in which the minimum (a) and (b) in Table \ref{tab:9Li} are adopted for $^9$Li, respectively.
For each calculation, we determine the parameter $\delta$ in the $^9$Li-$n$ potential, shown in Table~\ref{tab:energy_diff}.

\begin{table}[t]
\caption{
$\delta$ and the energy differences $\Delta E$ in MeV.}
\label{tab:energy_diff} 
\begin{center}
\begin{ruledtabular}
\begin{tabular}{c|cccc}
              &  Inert core   &  Pairing   &  Tensor   &  Present~~ \\
\noalign{\hrule height 0.5pt}
$\delta$        ~~  &  $0.066$ &  $0.143$  & $0.1502$   &  $0.1745$~~  \\
$\Delta E$      ~~  &    2.1   &  1.4      &  0.5       &  $-0.1$~~    \\
\end{tabular}
\end{ruledtabular}
\end{center}
\end{table}

\begin{figure}[t]
\centering
\includegraphics[width=8.2cm,clip]{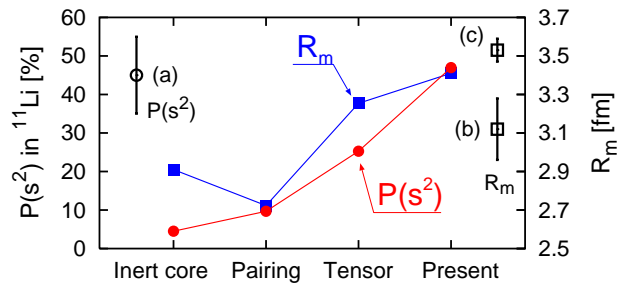}
\caption{(Color online) $(1s)^2$ probability $P(s^2)$ and matter radius $R_m$ of $^{11}$Li with four models in comparison with the experiments
((a)\cite{Si99}, (b)\cite{Ta88b} and (c)\cite{To97}). The scale of $P(s^2)~$($R_m$) is right (left) hand side.}
\label{11Li}
\end{figure}

In Fig.~\ref{11Li}, ``Present'' is found to give a large amount of the $(1s)^2$ probability $P(s^2)$, 46.9\% for the last two neutrons and a large matter radius $R_m$, 3.41 fm for $^{11}$Li, which are enough to explain the observations. The probabilities of $(p_{1/2})^2$, $(p_{3/2})^2$, $(d_{5/2})^2$ and $(d_{3/2})^2$ for the last two neutrons are obtained as $42.7\%$, $2.5\%$, $4.1\%$ and $1.9\%$, respectively. In Fig.~\ref{11Li}, when we individually consider the tensor and pairing correlations for $^9$Li, $P(s^2)$ is larger for the tensor case than for the pairing case. This means that the blocking effect from the tensor correlation is stronger than the pairing case. Finally, both blocking effects furthermore enhance $P(s^2)$ and provide almost equal amount of $(1s)^2$ and $(0p)^2$ configurations. Hence, two correlations play important roles to break the magicity and make the halo structure for $^{11}$Li.

In Table \ref{tab:energy_diff}, we also estimate the relative energy difference $\Delta E$ between $(1s)^2$ and $(0p)^2$ configurations for $^{11}$Li using the mixing probabilities of these configurations and the coupling matrix element between them as 0.5 MeV obtained in Ref.~\cite{My02}. The present model is found to give the degenerated energies enough to cause a large coupling between the $(0p)^2$ and $(1s)^2$ configurations by the pairing interaction between the last neutrons.

\begin{figure}[t]
\centering
\includegraphics[width=8.2cm,clip]{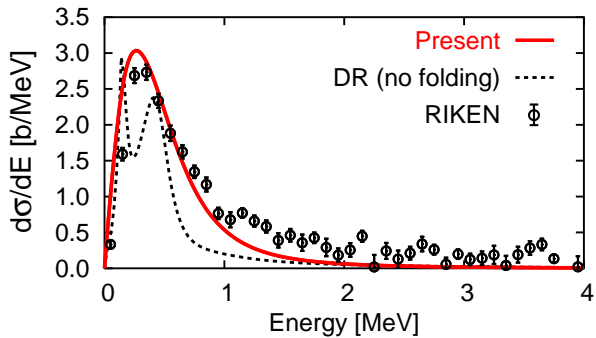}
\caption{(Color online) Calculated Coulomb breakup cross section measured from the $^9$Li+$n$+$n$ threshold energy.}
\label{cross}
\end{figure}

In addition to the matter radius, the halo structure also affects the proton radius of $^{11}$Li, 
because of the recoil effect of the c.m. motion. 
In the three-body model of $^{11}$Li, its proton radius ($R_p$) consisting of 
the proton radius of $^9$Li and the relative distance between $^9$Li and the c.m. of two neutrons ($R_{\rm c-2n}$)
with the following relation
\begin{eqnarray}
  \bra R^2_p(^{11}{\rm Li})\ket=\bra R^2_p(^9{\rm Li})\ket + \left(\frac{2}{11}\right)^2 \bra R_{\rm c-2n}^2\ket,
\end{eqnarray}
where the second term represents the recoil effect.
When the halo structure develops, $\bra R_{\rm c-2n}^2\ket$ is expected to be large.
Experimentally, considering the nucleon radius,
the charge radius of $^{11}$Li was measured recently and its value is 2.467$\pm$0.037 fm, which is
enhanced from the one of $^9$Li, 2.217$\pm$0.035 fm \cite{Sa06}.  
The improved calculation for the isotope shift determination\cite{Pu06} shows that 
2.423$\pm$0.037 fm and 2.185$\pm$0.033 fm for $^{11}$Li and $^9$Li, respectively.
The present wave functions provide 2.44 fm and 2.23 fm for $^{11}$Li and $^9$Li, respectively, 
which are in good agreement with the experimental values. 
This enhancement is mainly caused by the large value of $\sqrt{\bra R^2_{\rm c-2n}\ket}$ obtained as 5.69 fm.
For comparison, the distance between last two neutrons is 7.33 fm, which is larger than the core-$2n$ case.

We further calculate the three-body Coulomb breakup strength of $^{11}$Li into the $^9$Li+$n$+$n$ system to investigate the properties of the dipole excited states and compare the strength with the new data from the RIKEN group\cite{Na06}. We use the Green's function method combined with the complex scaling method\cite{Ao06} to calculate the three-body breakup strength\cite{My03} using the dipole strength and the equivalent photon method, where the experimental energy resolution is taken into account\cite{Na06}. We do not find any resonances with a sharp decay width enough to make a resonance structure. In Fig.~\ref{cross}, it is found that the present model well reproduces the experiment, in particular, for low energy enhancement and its magnitude.  Seeing more closely, however, our results seem to underestimate the cross section at $E>1$ MeV, while overestimate at low energy peak region slightly.
As a result, the integrated dipole strength for $E\leq 3$ MeV gives 1.35 $e^2 {\rm fm}^2$, which agrees with the experimental value of $1.42\pm0.18$ $e^2 {\rm fm}^2$\cite{Na06}.

For the reference, we calculate the strength with a potential model denoted as DR,
in which the $^9$Li core is inert and the $^9$Li-$n$ $s$-wave potential is deepened to reproduce $50\%$ of $P(s^2)$ 
in the $^{11}$Li ground state.
In this case, we obtain three dipole resonances of $1/2^+$, $3/2^+$ and $5/2^+$ states with $3/2^-\otimes1^-$,
less than 0.5 MeV above the three-body threshold energy, similar to the results of Ref.~\cite{Ga02b}.
In our results, the $3/2^+$ state is located slightly lower than other two states, 
because of the $J_0=0^-=(L=1)\otimes(S=1)$ component for two neutrons in Eq.~(\ref{WF11}),
where $L$ and $S$ are the coupled angular momenta and spins of the last two neutrons, respectively.
This component does not appear in the $1/2^+$ and $5/2^+$ states.
This difference for the dipole states makes a visible splitting 
in the cross section before folding with experimental resolution as shown in Fig.~\ref{cross}.
If we fold the spectrum by the experimental resolution, two peaks are washed out but
the strength distribution comes out different from both the present result and the experiment.
The detailed analysis of the dipole states would be shown in the forthcoming paper.

For $^{10}$Li, the present model successfully produces $-17.4$ fm for the scattering length of the $2^-$ state in the $^9$Li+$n$ system
as a signature of a virtual $s$-state\cite{Th99,Cha01,Je06}. The $1^-$ state gives $-5.6$ fm, not so large negative value.
Above the $^9$Li+$n$ threshold energy, two $p$-state resonances are obtained at 0.22 MeV and 0.64 MeV for $1^+$ and $2^+$ states 
with the decay widths of $0.09$ MeV and $0.45$ MeV, respectively.
From these results, it is found that the Pauli-blocking naturally describes the inversion phenomenon of $s$- and $p$-states in $^{10}$Li, in addition to the $^{11}$Li properties.


\section{Summary}

In summary, we have considered newly the tensor correlation in $^{11}$Li based on the extended three-body model.  We have found that the tensor and pairing correlations play important roles in $^9$Li with different spatial characteristics, where the tensor correlation prefers a shrunk spatial extension.
The tensor and pairing correlations in $^9$Li inside $^{11}$Li are then Pauli-blocked by additional two neutrons, which makes the $(1s)^2$ and $(0p)^2$ configurations close to each other and hence activates the pairing interaction to mix about equal amount of two configurations.
As a result we naturally explain the breaking of magicity and the halo formation for $^{11}$Li. 
We also reproduce the recent results of the Coulomb breakup strength and the charge radius of $^{11}$Li.
For $^{10}$Li, the inversion phenomenon is explained from the Pauli-blocking effect.

In this study, we focused on the tensor correlation, which is newly considered to figure out the $s$-$p$ shell gap problem.
However, the unified treatment of the effective interactions was not accomplished and is beyond the scope of this paper. This would require a consistent treatment of the short-range correlation in the realistic interaction while retaining the tensor force explicitly to describe the tensor correlation\cite{Fe98}.


\begin{acknowledgments}
The authors would like to thank Prof. I. Tanihata and Prof. H. Horiuchi for encouragement and Dr. S. Sugimoto and Prof. T. Nakamura for valuable discussions. 
This work was performed as a part of the ``Research Project for Study of
Unstable Nuclei from Nuclear Cluster Aspects (SUNNCA)'' at RIKEN and
supported by a Grant-in-Aid from the Japan Society for the Promotion of Science (JSPS, No. 18-8665).
Numerical calculations were performed on the computer system at RCNP.
\end{acknowledgments}

\end{document}